\documentclass{PoS}

\usepackage{graphicx}

\newcommand{\zaa}{\emph{Astron.~Astrophys.}}

\newcommand{\zapj}{\emph{Astrophys.~J.}}

\newcommand{\znp}{\emph{Nucl.~Phys.}}
\newcommand{\zpl}{\emph{Phys.~Lett.}}
\newcommand{\zpr}{\emph{Phys.~Rev.}}

\newcommand{\zadndt}{\emph{At. Data Nucl. Data Tables}}

\newcommand{\zjcap}{\emph{JCAP}}

\newcommand{\deu}{D}
\newcommand{\tro}{$^3$He}
\newcommand{\qua}{$^4$He}

\newcommand{\onz}{$^{11}$B}
\newcommand{\carb}{$^{12}$C}
\newcommand{\sep}{$^{7}$Li}
\newcommand{\hui}{$^{8}$Be}
\newcommand{\hli}{$^4$He, D, $^3$He and $^{7}$Li}
\newcommand{\aaag}{$^4$He($\alpha\alpha,\gamma)^{12}$C}
\newcommand{\aabe}{$^4$He($\alpha,\gamma)^{8}$Be}
\newcommand{\beac}{$^8$Be($\alpha,\gamma)^{12}$C}
\newcommand{\npd}{n(p,$\gamma$)d}
\newcommand{\hdp}{$^3$He(d,p)$^4$He}
\newcommand{\tdn}{$^3$H(d,n)$^4$He}

\newcommand{\gap}{\mathrel{ \rlap{\raise.5ex\hbox{$>$}}
                    {\lower.5ex\hbox{$\sim$}}  } }
\newcommand{\lap}{\mathrel{ \rlap{\raise.5ex\hbox{$<$}}
	            {\lower.5ex\hbox{$\sim$}}  } }

\title{Influence of the variation of fundamental constants on the primordial nucleosynthesis}

\ShortTitle{Influence of the variation of fundamental constants on the primordial nucleosynthesis}

\author{\speaker{Alain Coc}$^1$, Pierre Descouvemont$^2$,  Jean-Philippe Uzan$^3$ and Elisabeth Vangioni$^3$\\%
\llap{$^1$}
     Centre de Spectrom\'etrie Nucl\'eaire et de
Spectrom\'etrie de Masse (CSNSM), IN2P3-CNRS and Universit\'e Paris Sud 11, UMR~8609, B\^at. 104, 91405 0rsay
Campus (France)\\
\llap{$^2$}
Physique Nucl\'eaire Th\'eorique et Physique Math\'ematique, C.P. 229, Universit\'e Libre de Bruxelles (ULB), B-1050 Brussels, Belgium\\
\llap{$^3$}
Institut d'Astrophysique de Paris,
              UMR-7095 du CNRS, Universit\'e Pierre et Marie
              Curie,
              98 bis bd Arago, 75014 Paris (France)\\       
E-mail: \email{coc@csnsm.in2p3.fr, pdesc@ulb.ac.be, uzan@iap.fr, vangioni@iap.fr}}

\abstract{
We investigate the effect of a variation of fundamental constants on 
primordial element production in
Big Bang nucleosynthesis (BBN). We  focus on the effect of a possible 
change in the nucleon-nucleon interaction on nuclear reaction rates involving 
the $A=5$ ($^5$Li and $^5$He) and $A=8$ (\hui) unstable nuclei. 
The reaction rates for  \hdp\ and \tdn\ are dominated by the properties 
of broad analog resonances in $^5$Li and $^5$He compound nuclei
respectively.
While the triple--alpha process \aaag\ is normally not effective in BBN,  
its rate is very sensitive to the position of the ``Hoyle state" 
and could in principle be drastically affected if \hui\ were stable during BBN. 
We found that the effect of the variation of constants on the \hdp, \tdn\ and \aaag\ reaction rates is not sufficient to
induce a significant effect on BBN, even with a stable \hui. The main influences come 
from the weak rates and 
the $A=2$, n(p,$\gamma$)d, bottleneck reaction.
}

\FullConference{XII International Symposium on Nuclei in the Cosmos\\
		 August 5-12, 2012\\
		 Cairns, Australia}

\begin{document}

\section{Introduction}
\label{s:intro}

Constraints on the possible variation of fundamental constants are an efficient method of testing the equivalence principle~\cite{JPU11}, 
which underpins metric theories of gravity and in particular general relativity. 
These constraints are derived from a wide variety of physical systems and span a large time interval back to
Big Bang Nucleosynthesis (BBN).
Using inputs from WMAP for the baryon density~\cite{WMAP}, BBN yields excellent agreement between
the theoretical predictions and astrophysical determinations for the abundances of D and \qua~\cite{Cyb08b,CV10}
despite the discrepancy between the theoretical prediction of \sep\ and its determined abundance
in halo stars. 
The effects of the variation of fundamental constants on BBN predictions is difficult to model.  
However, one can proceed in a two step approach: first by determining the dependencies of 
the light element abundances on the nuclear parameters and then by relating those parameters
 to the fundamental constants, following our earlier work \cite{Coc07}.

It is well known that, in principle, the mass gaps at $A = $5 and $A=8$, prevent the nucleosynthetic
chain from extending beyond \qua. 
The presence of these gaps is caused by the instability of $^5$He, $^5$Li and \hui\
which are respectively unbound by 0.798, 1.69 and 0.092~MeV
with respect to neutron, proton and $\alpha$ particle 
emission.
Variations of constants will affect the energy levels of the $^5$He, $^5$Li, 
\hui\  and \carb\ nuclei \cite{Ber10,Eks10}, and hence, the resonance energies whose contributions
dominate the reaction rates.
In addition, since \hui\ is only slightly unbound, one can expect that 
for even a small change in the nuclear potential, it could become bound and may 
thus severely impact the results of Standard BBN (SBBN).
It has been suspected that stable \hui\ would trigger the production of heavy elements
in BBN, in particular that there would be significant leakage of the nucleosynthetic chain 
into carbon \cite{Bethe39}.  Indeed, as we have seen previously~\cite{Eks10}, changes in the nuclear potential
strongly affects the triple--alpha process and as a result, strongly affect the nuclear abundances
in stars. 

\section{Thermonuclear reaction rate variations}
\label{s:nucl}

It would be desirable to know the dependence of each of the main SBBN reaction rates to fundamental quantities. 
This was achieved in Ref.~\cite{Coc07}, but only for the first two BBN reactions: the n$\leftrightarrow$p weak interaction 
and the p(n,$\gamma$)d bottleneck. 
Here, we propose to extend this analysis to the \tdn\ and \hdp\ reactions that proceed through the $A=5$ 
compound nuclei $^5$He and $^5$Li, and to the \aaag\ reaction that could bridge the $A=8$ gap.   

The weak rates that exchange protons with neutrons can be calculated theoretically and their dependence 
on $G_F$ (the Fermi constant), $Q_{np}$ (the neutron--proton mass difference) and $m_e$ (the electron mass) is
explicit \cite{Dic82}. The dependence of the n+p$\rightarrow$d+$\gamma$ rate \cite{And06} cannot be directly related
to a few fundamental quantities as for the weak rates, but modeling of its dependence on the
binding energy of the deuteron $B_D$ has been proposed \cite{Dmi04,Car12}. 

For the \tdn, \hdp\ and \aaag\  reactions, we used a different approach.
In these three reactions, the rates are dominated by the contribution of  resonances  whose 
properties can be calculated within a microscopic cluster model. 
The nucleon-nucleon interaction $V(\mathbf{r})$ depends on the relative coordinate and is written as:
\begin{eqnarray}\label{eq2b}
 V(\mathbf{r})=V_C(\mathbf{r})+(1+\delta_{_{\rm NN}})V_N(\mathbf{r}),
\end{eqnarray}
where $V_C(\mathbf{r})$ is the Coulomb force and $V_N(\mathbf{r})$ the nuclear interaction. 
The parameter $\delta_{_{\rm NN}}$ characterizes the change in the  nucleon-nucleon interaction. 
When using the Minnesota force \cite{TLT77},
it is related to  the binding energy of deuterium by ${\Delta}B_D/B_D = 5.7701\times\delta_{_{\rm NN}}$ \cite{Eks10}.
(The variation of the Coulomb interaction is assumed to be negligible compared to the nuclear interaction).
The next important step is to relate $ {\Delta}B_D$ to the more fundamental parameters.
To summarize, $B_{\rm D}$ has been related, within an $\omega$ and $\sigma$ mesons exchange potential
to quark masses and $\Lambda_{\rm QCD}$ by Flambaum \& Shuryak \cite{Fla03} and subsequently to
more fundamental parameters (see Coc et al.~\cite{Coc07} and references therein), and in particular to the fine structure
constant.

\subsection{The triple--alpha}

The triple-alpha reaction is a two step process in which, first, two alpha--particles fuse into the \hui\ ground state,
so that an equilibrium (2$\alpha\leftrightarrow$\hui) is achieved.  
The second step is another alpha capture to the Hoyle state in \carb.
In our cluster approximation the wave functions of the \hui\ and \carb\ nuclei are approximated 
by two and three-cluster wave functions involving the alpha
particle, considered as a cluster of 4 nucleons.
It allows the calculation of
the variation of the \hui\ ground state and \carb\ Hoyle state w.r.t. the nucleon--nucleon interaction, i.e. $\delta_{_{\rm NN}}$.
In  Ref.~\cite{Eks10}, we obtained   
$E_{g.s.}(^8{\rm Be}) = \left( 0.09208-12.208\times\delta_{_{\rm NN}}\right)$ MeV, for the \hui\ g.s. and 
$E_R(^{12}{\rm C}) =  \left(0.2877-20.412\times\delta_{_{\rm NN}}\right)$ MeV, for the Hoyle state. 
From these relations, it is possible to calculate the partial widths, and subsequently the \aaag\ rate 
as a function of $\delta_{_{\rm NN}}$~\cite{Eks10}. Indeed, variations of $\delta_{_{\rm NN}}$ of the order of 1\%, induces orders of magnitude 
variations of the rate \cite{Eks10} at temperatures of a few 100 MK. 
In addition, one sees that $E_{g.s.}(^8{\rm Be})$ (relative to the 2--$\alpha$ threshold) becomes negative
(i.e. \hui\ becomes stable) for $\delta_{_{\rm NN}}\gap7.52\times10^{-3}$.
In that case,  we have to calculate the two reaction rates, \aabe\ and \beac\ for a stable \hui. 
The calculation of the rate of the second reaction can be achieved using the sharp resonance formula
with the varying parameters of the Hoyle state from Ref.~\cite{Eks10}. 
For the first reaction, \aabe,  we have performed a detailed calculation following \cite{Bay85} to
obtain the astrophysical $S$-factor, and reaction rate, 
for values of  the \hui\ binding energy of $B_8\equiv-E_{g.s.}(^8{\rm Be})$ = 10, 50 and 100 keV.

\subsection{The \hdp\ and \tdn\ reactions}

The \hdp\ and \tdn\ reactions proceeds through the $^5$Li and $^5$He compound nuclei and their 
rates are dominated by contributions of $\frac32^+$ analog resonances.
The corresponding levels are well approximated by cluster structures (\tro$\otimes$d or t$\otimes$d), so
that we can use the same microscopic model as for the \aaag\ reaction.   
However, unlike in the case of $^8$Be, the $^5$He and $^5$Li nuclei are unbound by $\sim$1~MeV 
and the resonances are broad.
Therefore the issue of producing $A=5$ bound states, or even  a two step process, like the triple--alpha reaction
is irrelevant.

\begin{figure}
\begin{center}
\includegraphics[width=.8\textwidth]{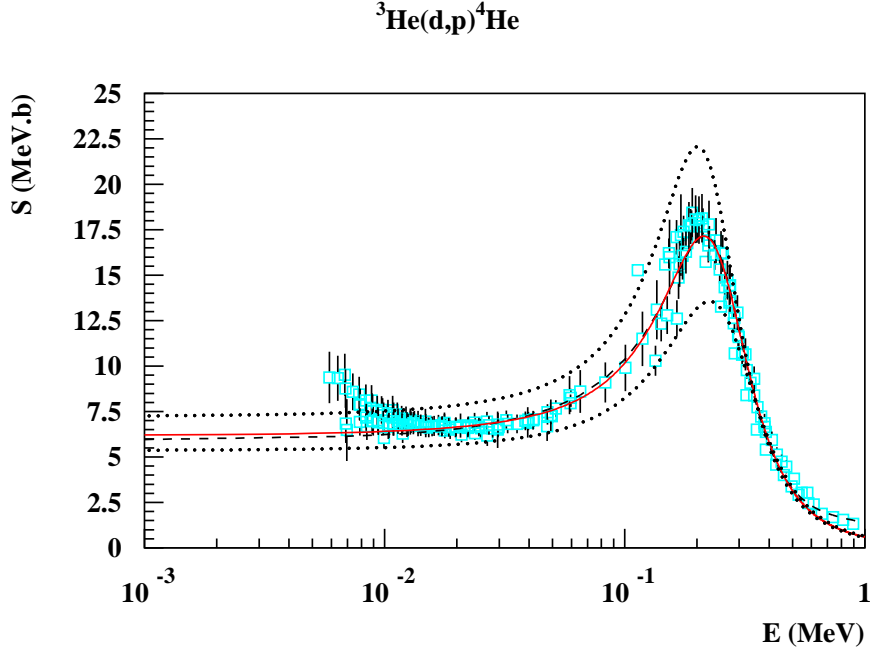}
\caption{$S$-factor curves for the \hdp\ reaction from Ref.~\cite{Des04} (dashed), from our fit (solid and overlapping) and for
extreme variations  of $\delta_{_{\rm NN}}$ =  $\pm$0.15 (dotted). 
Deviations with experimental data 
 at very low energy are due to screening. }
\label{f:hdp}
\end{center}
\end{figure}

To be consistent with our previous work, we want to reproduce, for $\delta_{_{\rm NN}}$=0, the  experimental $S$--factors 
(see references in Ref.~\cite{Des04})
obtained 
 by a full $R$--matrix analysis, but, here, for convenience, we restrict ourselves here to the single pole 
$R$--matrix approximation which will be shown to be sufficient: 
\begin{equation}
\sigma(E)\propto\frac{(\hbar c)^2}{\mu E}\frac{\Gamma_{\rm in}(E)\Gamma_{\rm out}(E)}{(E_R^*+\Delta E_R^*-E)^2 + \Gamma^{2}(E)/4}
\label{eq:sigma}
\end{equation}
For the \tdn\ reaction, we  use the parameterization of Barker~\cite{Bar97}, which reproduces the resonance 
corresponding to the $\frac{3}{2}^+$ state at 16.84 MeV which is in perfect agreement with the full $R$--matrix fit \cite{Des04} that 
we used in previous work \cite{CV10,Coc04,Coc09,Coc12a}. 
For the \hdp\ reaction, we performed a fit (see Fig.~\ref{f:hdp}) of the full R--matrix $S$--factor provided from Ref.~\cite{Des04} since, in that case,  
the parameterization of Barker~\cite{Bar97}, does not reproduce well the more recent data.

The additional level shifts obtained with our cluster model are  given by
${\Delta}E_R=-0.327\times\delta_{_{\rm NN}}$ for \tdn\ and ${\Delta}E_R=-0.453\times\delta_{_{\rm NN}}$
for \hdp\ (units are MeV) \cite{Coc12b}.  
These energy dependences are much weaker ($\sim$ 20--30~keV for $|\delta_{_{\rm NN}}|\leq0.03$) than
for $^8$Be and $^{12}$C. This is expected for broad resonances which are weakly sensitive to the nuclear interaction.
In contrast, Berengut {\em et al.}~\cite{Ber10} find a stronger energy dependence. 
These authors perform Variational Monte Carlo  calculations with realistic N-N interactions, which provide better D and $^3$H/$^3$He wave functions,
but which are not well adapted to broad resonances, such as those observed in $^5$He and $^5$Li. 
Besides, the calculation of ${\Delta}E_R$ as a function of $\delta_{_{\rm NN}}$ is obtained by the {\em difference} between the energy
of the ${3\over2}^+$ states and the thresholds for the two--clusters emission in the entrance channel, both depending on the
N--N-interaction.  
Berengut {\em et al.}~\cite{Ber10} assume that these levels follow the dependence of the $^5$Li and $^5$He ground 
states, but the ${3\over2}^+$ resonant levels state have indeed a \tro$\otimes$d or t$\otimes$d structure, different from the ground states. 
We also use a more elaborate parameterization of the cross--section.

\section{Effects on primordial nucleosynthesis}

\begin{figure}
\begin{center}
\includegraphics[width=.71\textwidth]{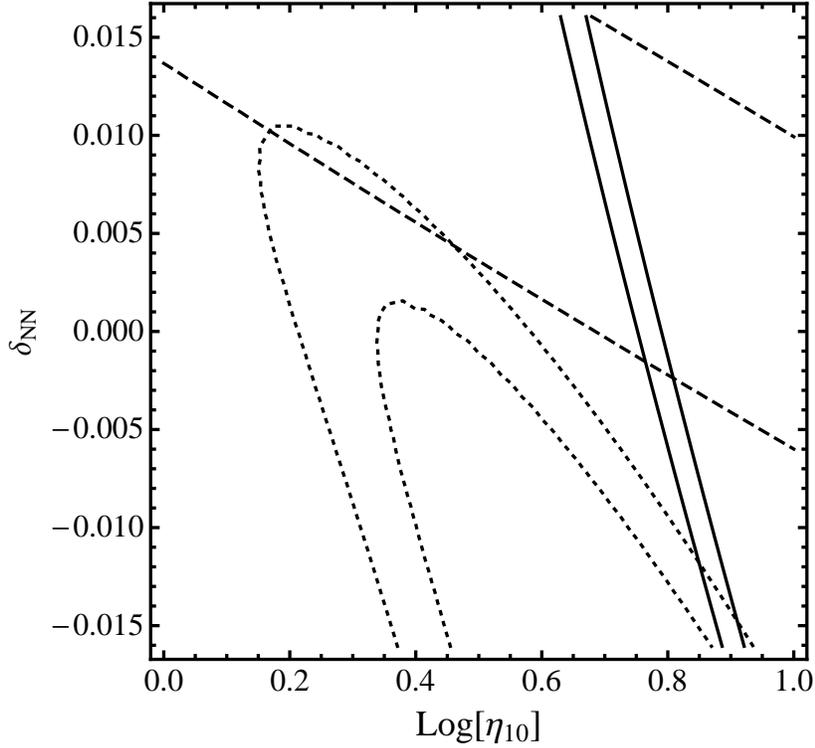}
\caption{Limits on  of $\eta_{10}$ (the number of baryons per 10$^{10}$ photons) and $\delta_{_{\rm NN}}$ provided by observational constraints 
on D (solid) \qua\ (dash) and \sep\ (dot).}
\label{f:contour}
\end{center}
\end{figure}

The results of the former sections can be implemented in a BBN code in order to compute the primordial abundances of the light elements as a 
function of $\delta_{_{\rm NN}}$. 
For \hli, we found that the effect of the \hdp\ and \tdn\ rate variations was negligible compared to the effect of the n$\leftrightarrow$p and 
n(p,$\gamma$)d reaction rate variations that we considered in our previous work \cite{Coc07}.   
Hence, next,  we allow those two last reaction rates to vary through the coupled variation of $\delta_{_{\rm NN}}$, $B_{\rm D}$,  
electron and quark masses, $G_F$,  $Q_{np}$, $\Lambda_{\rm QCD}$, etc.... as done in Ref.~\cite{Coc07}. 
Then, with updated \deu\ and \qua\ primordial abundances abundances deduced from observations, we obtained \cite{Coc12b}  
\begin{equation}
-0.0025 <\delta_{_{\rm NN}}< 0.0006.
\end{equation}
for typical values of the parameters. 
Those allowed variations in $\delta_{_{\rm NN}}$ are too small to reconcile \sep\ abundances with observations,
where $\delta_{_{\rm NN}}\approx-0.01$ is required. 
We can easily extend our analysis by allowing both $\eta_{10}$ and $\delta_{_{\rm NN}}$ to vary. 
This allows one to set a joint constraint on the two parameters $\delta_{_{\rm NN}}$ and baryonic density, as depicted on 
Figure~\ref{f:contour}. No combination of values allow for the simultaneous fulfilment of the \qua, \deu\ and \sep\ observational 
constraints.

Note that the most influential reaction on \sep\  is surprisingly \cite{Coc07,Fla03} n(p,$\gamma$)d as it affects
the neutron abundance and the $^7$Be destruction by neutron capture. The dependence of this rate to $B_{\rm D}$ that we used comes from 
Dmitriev et al. \cite{Dmi04} but, very recently, this has been challenged by the work of Carrillo et al. \cite{Car12} that provide a very different dependence.
If so, the influence of $\delta_{_{\rm NN}}$ on \sep\ would have to be re--evaluated.

Finally, we investigated the production of \carb\ by the \aaag, or the  \aabe\ and \beac\ reactions as a function of $\delta_{_{\rm NN}}$.
This is to be compared with the CNO (mostly \carb) SBBN production that has been calculated in a previous work \cite{Coc12a} to be 
CNO/H = $(0.5-3.)\times10^{-15}$, in number of atoms relative to hydrogen. A network of $\approx$400 reactions was used, but
the main nuclear path to CNO  was found to proceed from $^7$Li(n,$\gamma)^8$Li($\alpha$,n)$^{11}$B, followed
by    $^{11}$B(p,$\gamma$)$^{12}$C, $^{11}$B(d,n)$^{12}$C, $^{11}$B(d,p)$^{12}$B and 
$^{11}$B(n,$\gamma$)$^{12}$B reactions. 
To disentangle the \carb\ production through the \qua$\rightarrow$\hui$\rightarrow$\carb\ link, from the standard 
\sep$\rightarrow$$^8$Li$\rightarrow$\onz$\rightarrow$\carb\ paths,
we reduced the network to the reactions involved in $A<8$ plus the \aaag, or the  \aabe\ and \beac\ reactions,
depending whether or no \hui\ would be stable for a peculiar value of $\delta_{_{\rm NN}}$. 
The carbon abundance shows a maximum at $\delta_{_{\rm NN}}\approx0.006$, C/H$\approx10^{-21}$ \cite{Coc12b}, which is 
{\em six orders of magnitude} below the carbon abundance in SBBN \cite{Coc12a}. 
This can be understood as the baryon density during BBN remains in the range 10$^{-5}$ to 0.1 g/cm$^3$ between 
1.0 and 0.1~GK, substantially lower than in stars (e.g. 30 to 3000 g/cm$^3$ in stars considered by Ekstr{\"o}m et al.~\cite{Eks10}). This makes three-body reactions
 like \aaag\ much less efficient compared to  two-body reactions. 
 In addition, while stars can produce CNO at 0.1GK over billions of years, in BBN the optimal temperature
 range for producing CNO is passed through in a matter of minutes. 
 Finally, in stars, \aaag\ operates during the helium burning phase without significant sources of \sep, d, p and n to allow the 
 \sep$\rightarrow$$^8$Li$\rightarrow$\onz$\rightarrow$\carb, $A$=8, bypass process.

\begin{figure}
\begin{center}
\includegraphics[width=.8\textwidth]{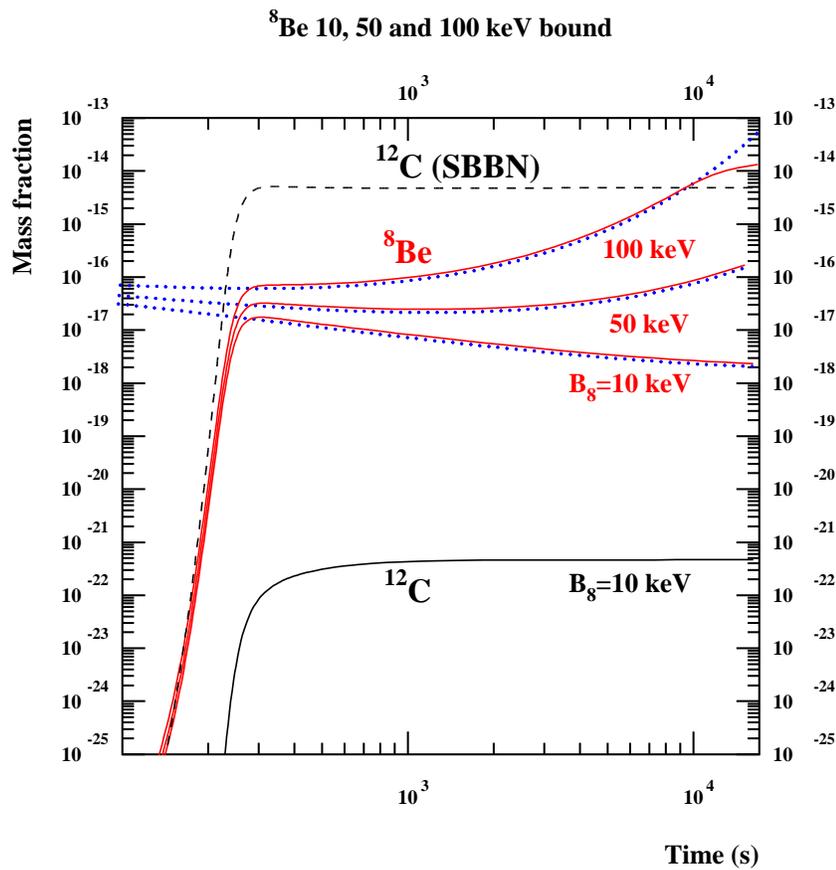}
\caption{\carb\ and \hui\ mass fractions as a function of time, assuming \hui\ is bound by 100, 50 and 10 keV 
 as shown by the upper to lower solid curves respectively \cite{Coc12b}.
 (Only the \carb\ mass fraction curve,  for $B_8$ = 10  keV, is shown; others
 are far below the scale shown). The dotted lines correspond to the computation at thermal equilibrium
 and the dashed line to the SBBN \cite{Coc12a} production.} 
\label{f:time}
\end{center}
\end{figure}

Note that the maximum is achieved for $\delta_{_{\rm NN}}\approx0.006$ when \hui\ is still unbound so that contrary to
a common belief, a stable \hui\ would not have allowed the buildup of heavy elements during BBN. This is illustrated
in  Figure~\ref{f:time} which displays the  evolution of the \carb\ and \hui\  mass fractions as a function of time when 
\hui\ is supposed to be bound by 10, 50 and 100~keV (solid lines).
They both increase with time until equilibrium between two $\alpha$--particle fusion and \hui\ photodissociation 
prevails as shown by the dotted lines. 
For the highest values of $B_8$, the \hui\ mass fraction increases until, due to the expansion,
equilibrium drops out, as shown by the late time behavior of the upper curve   ($B_8$ = 100 keV) in Figure~\ref{f:time}.
For $B_8\gap$ 10 keV, the \carb\ production falls well below, out of the frame, because the  \beac\ reaction rate
decreases dramatically due to the downward shift of the Hoyle state. 
For comparison, the SBBN $\approx$400 reactions network (essentially the \sep$\rightarrow$$^8$Li$\rightarrow$\onz$\rightarrow$\carb\ chain) result \cite{Coc12a}
is plotted (dashed line) in Figure~\ref{f:time}. It shows that for the  \qua$\rightarrow$\hui$\rightarrow$\carb\ path to give a significant contribution,
not only \hui\ should have been bound by much more than 100~keV, but also the \beac\ rate should have been much
higher in order to transform most \hui\ in \carb.

\section{Discussion}\label{sec5}

We have investigated the influence of the variation of the fundamental constants on the predictions of BBN and extended
our previous analysis~\cite{Coc07}.
Through our detailed modeling  of the cross-sections we have shown that, although the variation of the nucleon-nucleon potential
can greatly affect the triple--$\alpha$ process, its effect on BBN and the production
of heavier elements such as CNO is typically 6 orders of magnitude smaller than standard model abundances.  
Even when including the possibility that $^8$Be can be bound, at the temperatures, densities and 
timescales associated with BBN, the changes in the \aaag\ and \beac\ reaction rates
are not sufficient.
We have also extended our previous analysis by including effects involving $^5$He and $^5$Li. 
This allowed us to revisit the constraints obtained in Ref.~\cite{Coc07} and in particular to show that the effect of the 
\hdp\ and \tdn\ cross-sections variations remain small compared to the \npd\ induced variation. 

\acknowledgments
 
This work was partly supported by the French ANR VACOUL.

\end{document}